\documentclass[superscriptaddress,twocolumn,prb,showpacs]{revtex4}
\usepackage{graphicx}
\usepackage{dcolumn}
\usepackage{bm}
\usepackage{amsmath}
\usepackage{amssymb}
\usepackage{color}

\begin{document}

\title{Size Effects on Transport Properties in Topological Anderson
Insulators}
\author{Wei Li}
\author{Jiadong Zang}
\affiliation{Department of Physics, Fudan University, Shanghai,
200433, People's Republic of China}
\author{Yongjin Jiang}
\affiliation{Center for Statistical and Theoretical Condensed Matter
Physics, and Department of Physics, Zhejiang Normal University,
Jinhua 321004, People's Republic of China}

\date{\today}

\begin{abstract}
We study the size effects on the transport properties in topological Anderson insulators by means of the
Landauer-B\"{u}ttiker formalism combined with the nonequilibrium Green function method. Conductances calculated for serval different widths of the nanoribbons reveal that there is no longer quantized plateaus for narrow nanoribbons. The local spin-resolved current distribution demonstrates that the edge states on the two sides can be coupled, leading to enhancement of backscattering as the width of the nanoribbon decreases, thus destroying the perfect quantization phenomena in the topological Anderson insulator. We also show that the main contribution to the nonquantized conductance also comes from edge states. Experiment proposals on topological Anderson insulator are discussed finally.
\end{abstract}

 \maketitle

Recently, the study of the topological insulator (TI) has triggered great research interests\cite{CLKane,Qi_2010,Qi_2011}.
It was first proposed both from topological band
theory and topological field theory, and soon experimentally
realized in HgTe/CdTe quantum wells (QWs)\cite{Bernevig_2006,
Konig_2007}. Although it is insulating in the bulk, conduction is
allowed on the boundary due to the presence of edge states. These
edge states are protected by the time reversal symmetry. Each edge state is accompanied by its time
reversal partner, and the number of pairs is an odd number for
topological nontrivial phase, which leads to an odd-integer
quantized conductance on each edge. However these stories are
restricted to where the TI is semi-infinite. When two edges are
getting close to each other, finite size effect\cite{BinZhou,Linder} plays an important role. The overlap
between edge states from opposite edges opens an energy gap so
that properties of TI can be subtly modified.

The concept of TI can be generalized into many
other insulators, such as topological Mott insulator\cite{Raghu}, topological
superconductor\cite{Fu_TSC_2008,Fu_TSC_2009,Fu_TSC_PRL}. Its generalization in Anderson insulator (AI) is also addressed in disordered system%
\cite{JianLi2009,QFSun2009,Beenakker2009,HMGuo2010} recently. It is
believed that the disorder-induced edge states are topologically
protected and play a central role. In this sense such insulator is
named as topological AI (TAI). Thus, a question
arises naturally that whether this phenomenon will also be
emergent in the narrow nanoribbon with finite size. It is
interesting to explore the role edge state plays in the finite
size system.

In the present paper, we studied size effects on TAI in detail. We choose
to study the typical topological band insulator formed by
HgTe/CdTe QWs\cite{Bernevig_2006}, where spin-orbit
coupling is encoded. TAI phase is addressed at certain random
strength when the Fermi surface is at the bulk conduction band,
where a quantized conductance is observed. We show that the
conductance is no longer quantized plateaus for narrow
nanoribbons. To understand clearly the physics of this picture, we
presented the local spin-resolved current distribution in the
disordered bar, which demonstrated that the main contribution to
the nonquantized conductance comes from edge states. However, due
to the truncation of the coherence length between two edges by the
finite sample size, disorder can induce the interedge scattering.
As a result, the TAI phase will be suppressed. Through detailed
size dependence study, we found that by decreasing the width of
the nanoribbon, the coupling between edge states will lead to
exponential enhancement of the backscattering which destroys the
TAI phase eventually.

As a starting point, we consider a HgTe/CdTe QWs narrow nanoribbon.
The low energy electron states are approximately described by an
effective four band Hamiltonian\cite{Bernevig_2006}:

\begin{equation}
{\hat{\mathcal{H}}}=\left(
\begin{array}{cc}
h(\mathbf{k}) & 0 \\
0 & h^{\ast }(-\mathbf{k})%
\end{array}%
\right)  \label{eq:one}
\end{equation}%
where $h(\mathbf{k})=\epsilon (\mathbf{k})+\vec{d}(\mathbf{k})\cdot \hat{\sigma} $, $\mathbf{k}=(k_{x},k_{y})$ is
the two-dimensional wave vector, $\hat{\sigma} =(\hat{\sigma}_{x},\hat{\sigma}_{y},\hat{\sigma}_{z})$ are Pauli matrices. Up to the lowest order of
$\mathbf{k}$, $\vec{d}(\mathbf{k})=(Ak_{x},Ak_{y},M-Bk^{2})$, and $\epsilon (\mathbf{k})=C-Dk^{2}$,
where the parameters A, B, C and D depend on the thickness of
HgTe/CdTe QWs. $h^{\ast }(-\mathbf{k})$ is nothing but the time reversal
counterpart of $h(\mathbf{k})$ so that time reversal symmetry is respected.
The Hamiltonian is obtained by reducing the
eight-band Kane model to the reduced Hilbert space $|E1,1/2>$, $|H1,3/2>$, $%
|E1,-1/2>$, and $|H1,-3/2>$. Mass $M$ is an important parameter
describing
the energy gap between conduction and valence band. The Hamiltonian with $%
M>0 $ describes a conventional band insulator, while $M<0$
corresponds to the TI. In the present study,
tight-binding lattice model is used, so that the above effective
model is compactified by
substitutions $k_{i}\rightarrow \frac{1}{a}\sin(k_{i}a)$, and $%
k_{i}^{2}\rightarrow \frac{2}{a^{2}}(1-\cos(k_{i}a))$, where
$i=x,y$, and $a$ is the lattice constant. The width and length of
the nanoribbon under study are $L_{y}$ and $L_{x}$ respectively.
Meanwhile, we introduced disorders
through random on-site energy with a uniform distribution within $[-W/2,W/2]$%
, with the disorder strength $W$.

\begin{figure}[tbp]
\includegraphics[bb=12 12 323 245,width=7.5cm,height=6.5cm]{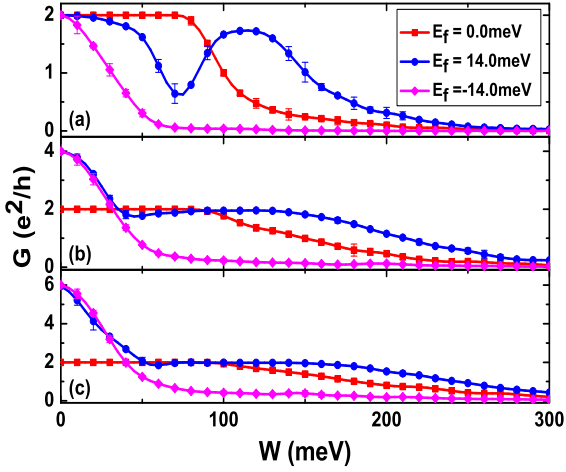}
\caption{(Color online) The conductance $G$ (a)-(c) vs disorder strength $W$ for different
Fermi energy $E_f$. The widths $L_{y}$ of the nanoribbon are (a) $L_{y} = 100nm$, (b) $L_{y} =
200nm$, and (c) $L_{y} = 300nm$.} \label{fig:fig1}
\end{figure}

To calculate the transport properties, we applied the method by
nonequilibrium Green function\cite{QFSun2009}. A small external bias $%
V=V_{L}-V_{R}$ is applied longitudinally between the two terminals.
The local current between neighboring sites $i$ and $j$ is
calculated by the formula\cite{Hatsugai_PRB_1993,Jauho,Nakanishi,YanyangZhang}:

\begin{equation}
J_{i\rightarrow j}=\frac{2e^{2}}{h}Im[\sum_{\alpha\beta}{\hat{\mathcal{H}}}%
_{i\alpha,j\beta}G_{i\beta,j\alpha}^{n}(E_{f})](V_{L}-V_{R})]
\label{eq:three}
\end{equation}%
where $V_{L(R)}$ describes the voltages at the lead-L(R). $%
G^{n}(E_{f})=G^{r}\Gamma _{L}G^{a}$ is electron correlation function
with
line width function $\Gamma _{L(R)}=i[\Sigma _{L(R)}^{r}-\Sigma _{L(R)}^{a}]$%
, and the retarded Green functions
$G^{r}(E_{f})=[G^{a}(E_{f})]^{\dag }=1/[E_{f}-H_{cen}-\Sigma
_{L}^{r}-\Sigma _{R}^{r}]$, with $H_{cen}$ the Hamiltonian in the
central region. The local spin-resolved current
$J^\alpha_{i\rightarrow j}$ between neighboring sites $i$ and $j$
with spin index $\alpha$ can also be calculated from Eq. (\ref{eq:three}) without summing over spin index $\alpha$. The net current $J_{L}$ flowing
through the central region is calculated by summing index $i$ for
local currents $J_{i\rightarrow i+\hat{x}}$ over any cross-section. After obtaining the current $%
J_{L} $, the linear conductance is given by $G=\lim_{V\rightarrow 0}dJ_{L}/dV$%
. In addition, the linear conductance can be directly obtained by $%
G=Tr[\Gamma _{L}G^{r}\Gamma _{R}G^{a}]$.

In the following numerical calculations, we used the realistic
material parameters of the HgTe/CdTe QWs\cite{Konig_2007}: $A=364.5$meV$\cdot $nm, $B=-686$%
meV$\cdot $nm$^{2}$, $C=0$meV, $D=-512$meV$\cdot $nm$^{2}$, and
$M=-10$meV.
The length of the nanoribbon $L_{x}=1000$nm, and the lattice constant $a=5$%
nm. Since this model is only valid for small $k$, we set the Fermi
energy small around the $\Gamma $ point. In the presence of
disorder, the
conductance is averaged over up to 400 random configurations except for Fig. %
\ref{fig:fig4} where 800 random configurations are used for each
data.

\begin{figure}[tbp]
\includegraphics[bb=10 10 300 230,width=7.5cm,height=5.0cm]{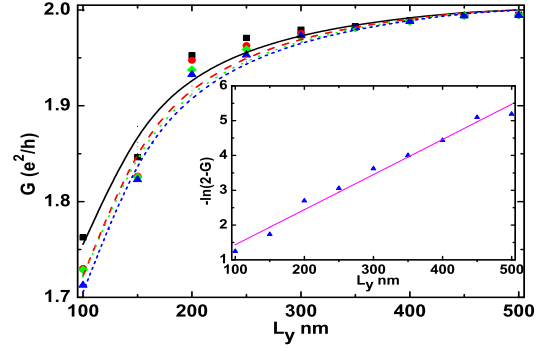}
\caption{(Color online) The conductance $G$ vs bar width $L_y$ at disorder
strength $W=115m$eV. Different lengths are chosen with $L_{x}$=1000nm (black solid line), 2500nm (red dash line), 4000nm (green dot line), and 5000nm (blue short dash line), while the symbols {\tiny\textcolor{black}{$\blacksquare$}}, {\tiny\textcolor{red}\textbullet}, {\tiny\textcolor{green}{$\blacklozenge$}} and {\tiny\textcolor{blue}{$\blacktriangle$}} are corresponding to the
realistic results of exact diagonalization, respectively. The inset shows the linear fitting of ln(2-G) by picking a
typical length $L_x=5000nm$.} \label{fig:fig2}
\end{figure}

We first studied the conductance $G$ versus disorder strength $W$
for different Fermi energy $E_{f}$ and the width $L_{y}$ of the
nanoribbon, as shown in Fig.\ref{fig:fig1}. As long as the Fermi
surface $E_{f}=0.0$meV (see Fig.\ref{fig:fig1}) lies inside the bulk gap, the conductance
maintains a quantized value $2e^{2}/h$ for different widths in the
clean limit. The conductance remains this quantized value in a
broad range of the disorder. When the disorder further increases,
electronic states become localized, and the conductance decreases
to zero rapidly. Such observation agrees well with the previous
result that the quantum spin Hall effect\cite{Kane1,Kane2,LFu,LSheng} is robust against weak
disorder and independent of the width of the nanoribbon. It's well
known that a finite size of the Hall bar opens a gap on the edge
due to overlap between edge states\cite{BinZhou}. Thus the edge
states are no longer Dirac particles. Our result shows that
although the edge states gain mass in nanoribbon, the topology is
still preserved. Backscattering is still forbidden due to time
reversal symmetry according to a brilliant argument in Ref.
\onlinecite{Qi_2010}.

\begin{figure}[tbp]
\begin{center}
\tabcolsep=0cm
\begin{tabular}{cc}
\includegraphics[bb=100 282 480 540,width=7.5cm,height=3.1cm]{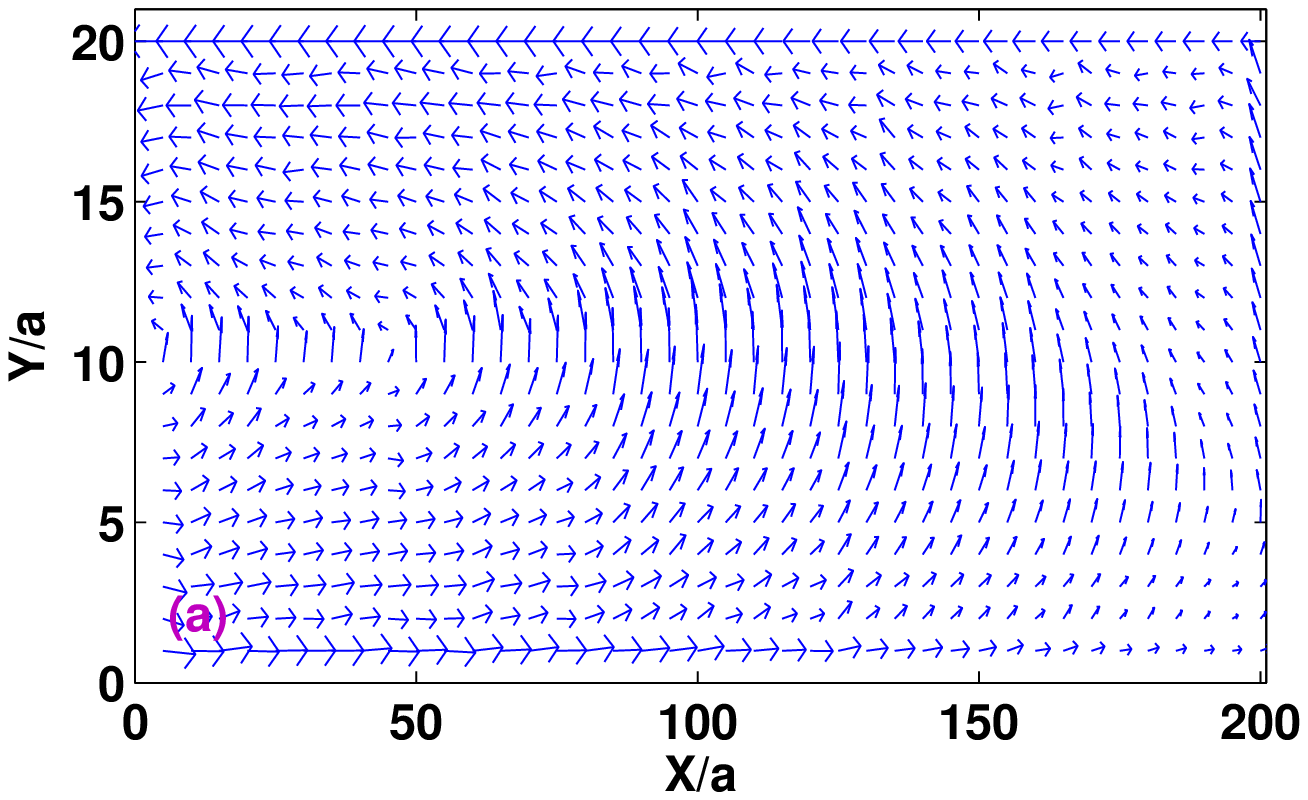} &
\\
\includegraphics[bb=100 282 480 540,width=7.5cm,height=3.1cm]{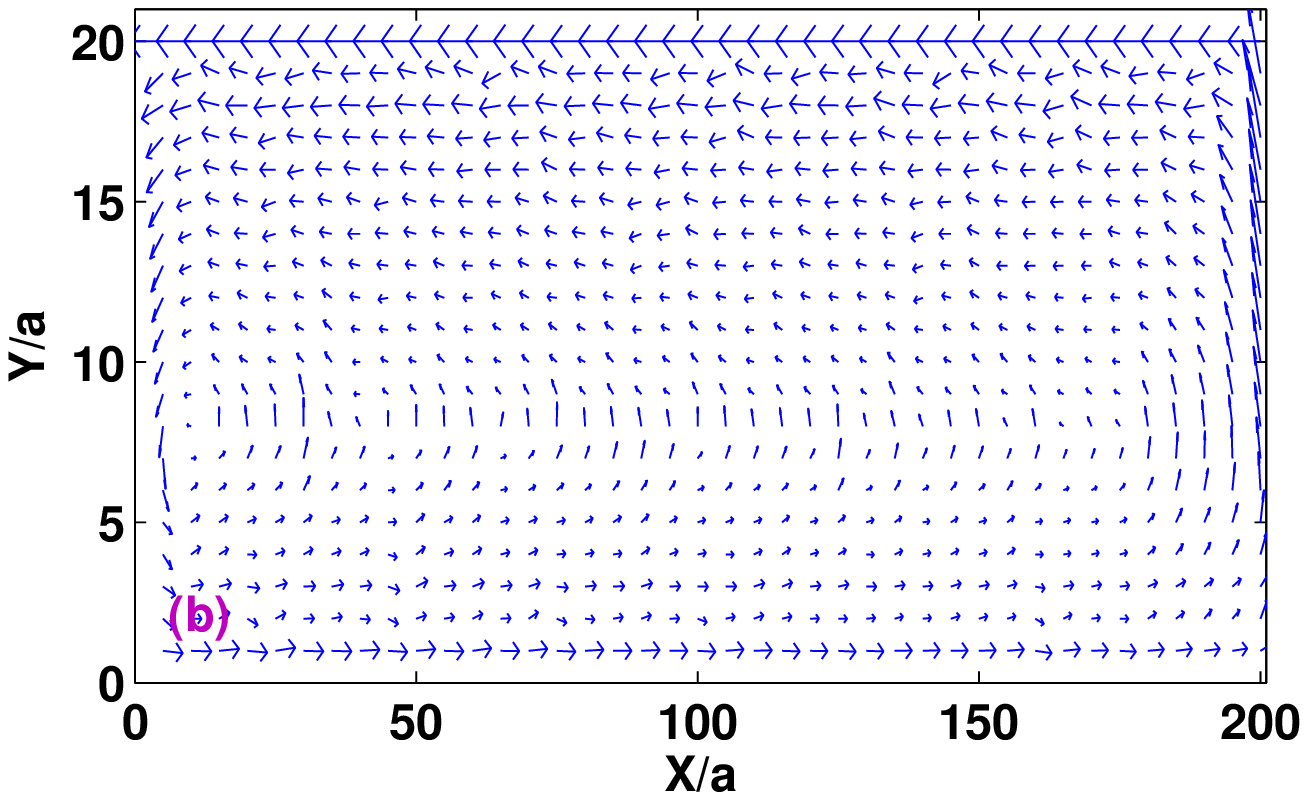} &
\\
\includegraphics[bb=100 282 480 540,width=7.5cm,height=3.1cm]{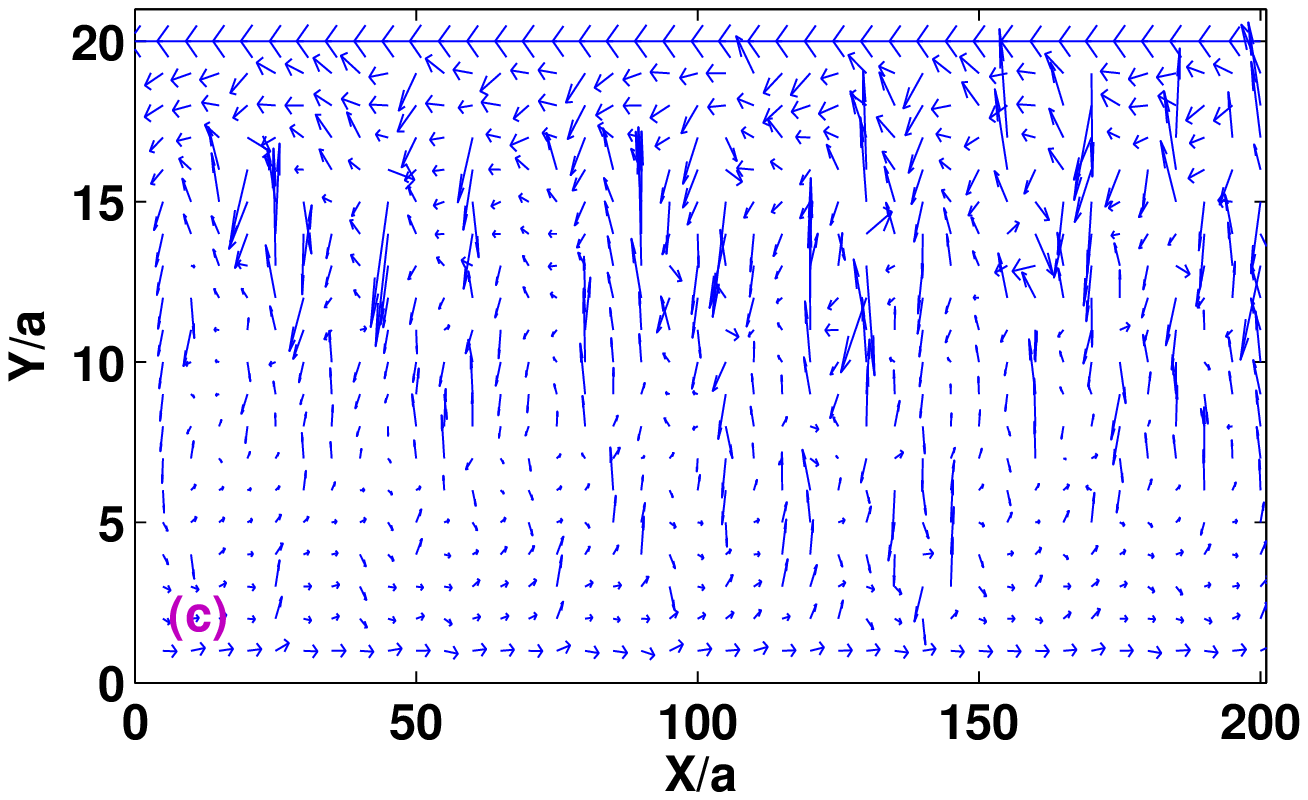} &
\end{tabular}%
\end{center}
\caption{(Color online) The distribution of local spin-resolved current with disorder strength (a) $W = 75.0m$eV, (b) $W = 90.0m$eV
, and (c) $W = 115m$eV. The Fermi energy is $E_{f} = 14.0m$eV, and
the bar size is $L_{x} = 200a$, and $L_{y} = 20a$. The vector length
is proportional to the square root of the current value.}
\label{fig:fig3}
\end{figure}

When the Fermi surface lies in the valence band of the bulk such
as $E_{f}=-14$meV shown in Fig. \ref{fig:fig1}, system is metallic
in the absence of disorder. Once the disorder is turned on,
electrons become localized, and the conductance dives to zero
directly (see Fig. \ref{fig:fig1}). This result exhibits the
conventional Anderson localization phenomenon. On the other hand,
when the Fermi energy $E_{f}$ is raised up into the bulk conduction band, say $E_{f}=14$%
meV, and the width is sufficiently large [see Fig. \ref{fig:fig1}(b) and (c)], conductance $G$ decreases gradually when
the strength of the disorder increases. Comparing to the Fig.
\ref{fig:fig1}(b) and (c), it is apparent that the
\textquotedblleft dip" in conductance (which occurs at $W\sim
60meV$) becomes less and less pronounced as the nanoribbon width
$L_y$ increases. Because the increasing of the nanoribbon width
$L_{y}$, the number of conducting channels in the bulk is
increased. The more conducting channels there are, the larger the
conductance will be.
In addition, the relation between conductance $G$ and
the nanoribbon width $L_{y}$ obeys: $GL_x=\sigma_c L_y$, where
$\sigma_c$ is the conductivity, which is width independent.
Beyond a certain onset of $W$, the conductance turns back and
increases to an approximate quantized value ($2e^{2}/h$). $G$
maintains this value for a certain range of $W$ before eventually
decreases. Concerning the spin degeneracy, each spin component
contributes one conductance quanta only. This odd-integer valued
conductance plateau indicates that the the system is in a
topologically distinct phase on this stage. More importantly,
different from the pure TI, the quantized value is induced by
disorder. This topological nontrivial phase is exactly the TAI
phase. However, one should note that
when the width is as small as $L_{y}=100$nm, as shown in Fig. \ref{fig:fig1}%
(a), although the plateau is still present, it is no longer
integer quantized, but irrationally valued in unit of $e^{2}/h$.
At the same time, the conductance plateau evolves into a hump
structure.

To further understand the finite size effect on TAI, the relation
between hump peak value and bar size is investigated in detail, as
shown in Fig. \ref{fig:fig2}. With the increase of width, the
plateau conductance $G$ increases as well, and finally saturates
to $2e^2/h$. After numerous calculation, we found that the
quantity $(2-G)$ behaves as an exponential function of the width
$L_y$, which is quite unconventional. In the case of normal metal,
conductance is proportional to the width due to the increase of
channels. This exponential behavior of conductance suggests the
conductance hump is contributed also by the edge channel of the
Hall bar, as the interaction between states localized on opposite
edges decays exponentially with respect to the width.
Due to previous studies by Groth \textit{et al}
\cite{Beenakker2009}, finite disorder would reverse the sign of
mass term of the BHZ model\cite{Bernevig_2006}, leading to a
robust edge state. The above result adds new insights to this
understanding in narrow nanoribbon structure.

\begin{figure}[tbp]
\includegraphics[bb=10 10 750 550,width=8.5cm,height=7.5cm]{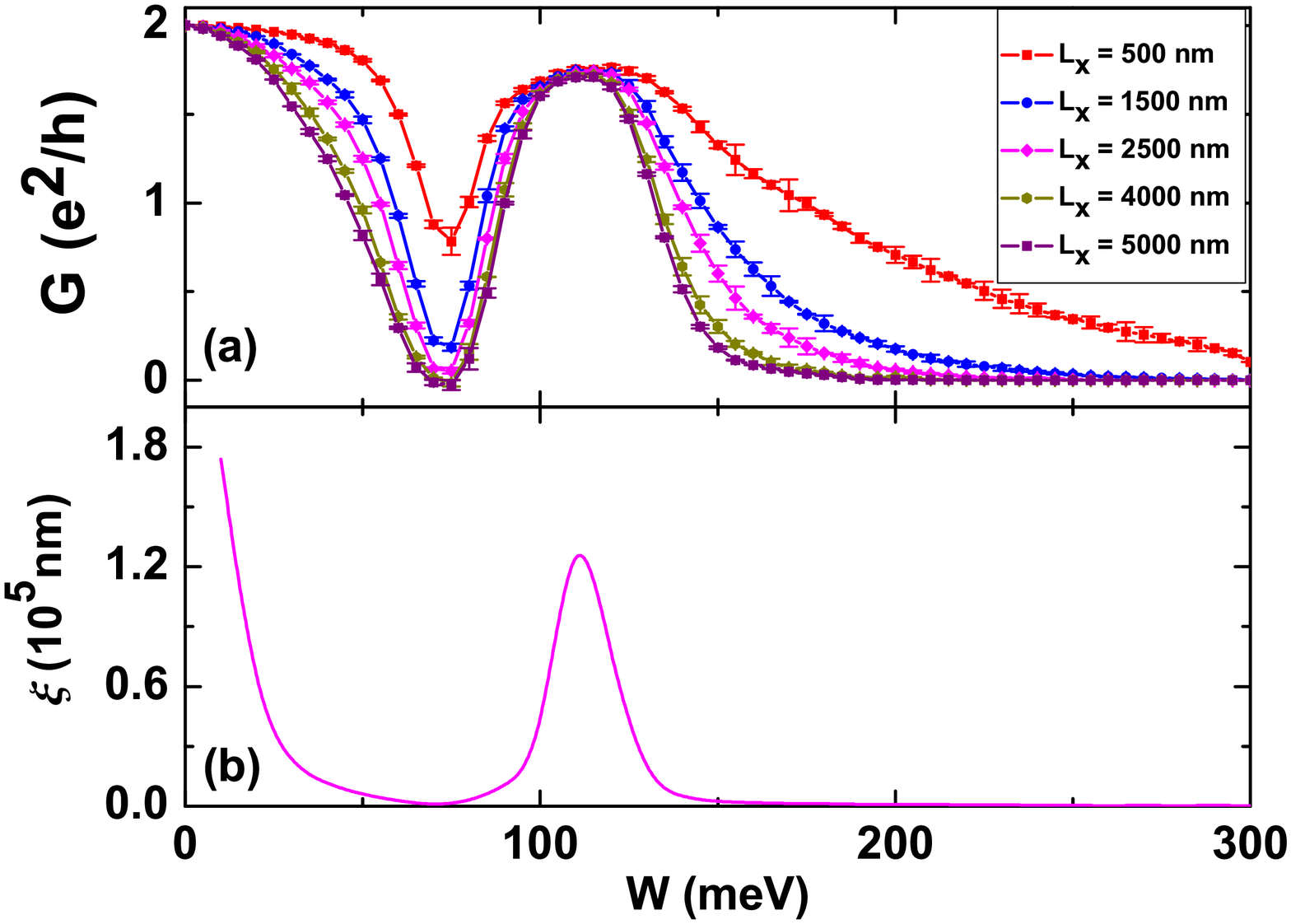}
\caption{(Color online) (a) the conductance $G$ vs disorder strength $W$ for different lengths of the nanoribbon. (b)
the localization length $\xi$ vs disorder strength $W$. The Fermi energy $E_{f} = 14.0m$eV and the width of the
nanoribbon is $L_{y} = 100nm$.} \label{fig:fig4}
\end{figure}

In order to get a better insight into the microscopic origin of
the conductance variations as shown in Fig.
\ref{fig:fig1}(a), the local spin-resolved current distributions
are plotted in Fig. \ref{fig:fig3}. Due to the time reversal
symmetry, we only consider the spin-up component, the influence of
spin-down component can be directly obtained by time reversal
symmetry. Here, the local spin polarized current on site $i$ is
defined as $J^{\uparrow}_{i} = J^{\uparrow}_{i\rightarrow
i+\hat{x}} + J^{\uparrow}_{i\rightarrow i+\hat{y}}$. When $W=75.0$meV, the whole sample behaves similarly
as a conventional metal with large backscattering. The vortex-like
circulation pattern can be easily seen, which reflects that the
scattering direction are determined by the chirality of spin
orbital interaction. As to the transport property, this is an
AI as the conductance decreases very rapidly as an
exponential form with respect to the bar length $L_x$ [see
Fig.4(b)]. As the disorder strength increases gradually, current
on one edge is greatly suppressed, while enhanced on the other
edge. Meanwhile, the bulk of the sample becomes more and more
insulating. All these signals indicates a formation of stable edge
states. At last, when the disorder is raised up to $W=115.0$meV
and conductance reaches peak value, bulk states are extremely
disordered and insulating. Edge states on the upper edge survives
and appears to be very robust against disorder.  However, due to
the finiteness of the nanoribbon width, small backscattering is still
present by hopping from the upper edge to the lower one. The
reduction of conductance is apparently proportional to this
hopping probability which is an exponential function of the width.

In Fig. \ref{fig:fig4}(a), the conductance $G$ as a function of
disorder strength $W$ with different length $L_x$ of the
nanoribbon are plotted. The dip feature is clearer for a large
$L_x$ because of the increasing probability of the backscattering
between the two edges. It is interesting to point out that the
main shape of the conductance hump at around the disorder strength
$W=115$meV changes little with the bar length, and the system
in that region exhibits the nonquantized version of TAI behavior. The corresponding
localization length $\xi = -lim_{L_x\rightarrow\infty}L_x/$ln$(G)$
as a function of disorder strength $W$ is shown in Fig.
\ref{fig:fig4}(b). We notice that the localization length
increases dramatically near the conductance hump region, which is
the finite size version of the picture that reasonable disorder
scattering can drive an insulating system into a TAI phase so that
the system shows quantized
conductance\cite{JianLi2009,Beenakker2009}. On the other hand,
$\xi$ almost vanishes near the dip region. From the theoretical
viewpoint, the sharp difference of longitudinal localization
length in different disorder regions plays the central central
role for all experimental observable effects, both for bulk and
finite size version of  TAI.

In summary, the size effects on the transport properties in TAI are
studied in this paper. We found the
conductance plateau deviates from its
quantized value exponentially as the nanoribbon width decreases. Such
behavior is originated from the mutual interaction between edge
states, as is clearly evidenced by local spin-resolved current
distributions in the disordered bar. Furthermore, we found the
conductance is generally decaying exponentially as the system
length increases. However, in the TAI region, the longitudinal
localization length is extremely long, which is found up to
hundreds of microns. This is quite a good feature that makes the
TAI phase with a narrow nanoribbon experimentally
distinguishable.

It is reported\cite{BBeri} that topological insulator can be realized in optical lattices, where the disorder can be introduced by an optical laser speckle potential\cite{JELye,Billy}. By tuning the laser intensity, it's quite promising to observe the TAI and the finite size effect. However in contrast to the uniform random disorder studied in this work, the disorder in optical laser speckle is correlated. The effect of this difference is unknown up to now. Therefore, it should
be an interesting problem for future studies.


We gratefully acknowledge the financial support from the National
Natural Science Foundation of China under grant No.11004174. WL and JZ thank the support by Fudan Research Program on Postgraduates. WL thanks Qing-feng Sun, and Jian Li for helpful discussions.

\end{document}